\documentclass{ws-p8-50x6-00}
\usepackage{amsmath}
\usepackage{epsfig}
\begin{document}
\title{Centrality Dependence of Baryon
and Meson Momentum Distributions in $pA$
Collisions\footnote{Talk presented in the 31th International Symposium on Multiparticle
Dynamics, Datong, China}}
\author{C.\ B.\ Yang$^{1,2}$  and  Rudolph C. Hwa$^2$}
\address{$^1$
Institute of Particle Physics, Hua-Zhong Normal
University,\\ Wuhan 430079, P.\ R.\ China\\
$^2$Institute of Theoretical Science and Department of
Physics, \\ University of Oregon, Eugene, OR 97403-5203, USA}
\maketitle

\abstracts{The proton and neutron inclusive distributions in the
projectile fragmentation region of $pA$ collisions are
studied in the valon model.  Momentum degradation
and flavor changes due to the nuclear medium are
described at the valon level using two parameters.
Particle production is treated by means of the
recombination subprocess.   Pion
inclusive distributions can be calculated without any
adjustable parameters.\\
PACS Number(s): 25.75.Dw, 13.85.Ni}

The study of proton-nucleus $(pA)$ collisions is important
because they are tractable intermediaries between $pp$ and
$AA$ collisions, when intense interest exists in discovering
the extent to which the dense medium created in an $AA$
collision differ from that of linear superpositions of $pp$
collisions.  One of the properties of $pA$ collisions is that the
momentum of the leading baryon in the projectile
fragmentation region is degraded, a phenomenon
commonly referred to, somewhat inappropriately, as baryon
stopping.  Such a transference of baryon number from the
fragmentation to the central region contributes to the
increase of matter density at mid-rapidity, thereby raising
the likelihood of the formation of quark-gluon plasma.  Thus
it is important to understand the process of baryon
momentum degradation  and its dependence on
nuclear size or centrality.

Since it is not feasible to perform first-principle
parameter-free calculations of the momentum degradation at this
point, experimental guidance is of crucial importance.
Recently, several experiments have produced useful data on
the subject, in particular, E910 and E941 at the AGS,
and NA49 at the SPS \cite{bc}.  It is the $x_F$
dependences of the distributions of $p-\bar{p}$ and
$n-\bar{n}$ that we shall focus on; moreover, their
dependences on centrality will guide us in our
determination of the nuclear effect on baryon momentum degradation.

 Since it is questionable that the concept
of color strings can be relevant in heavy-ion collisions where
the abundance of color charges in the overlap region renders
unlikely the development of constricted color flux tubes
\cite{rch2}, our approach in this paper will be on the various
levels of the constituents of the nucleon that are consistent
with the parton model.  More specifically, we shall use the
valon model \cite{rch3}-\cite{rch5} to keep track of the
momenta of the constituents and the recombination model
\cite{rch5,dh} to describe the hadronization of the partons.
The valons play a role in the collision problem as the
constituent quarks do in the bound-state problem.  Thus a
nucleon has three valons which carry all the momentum of
the nucleon, while each valon has one valence quark and its
own sea quarks and gluons.  Although soft processes are
non-perturbative, the valon model (including
recombination) nevertheless provides a systematic way of
calculating all subprocesses that contribute to a particular
inclusive process.  Some of the subprocesses can be
identified with certain diagrams in other approaches, e.g.,
baryon junction and diquark breaking terms \cite{dk,vgw}.
The valon distributions in the proton will be determined by
fitting the parton distributions at low $Q^2$.  The effect of
the nuclear medium on the valon distribution of the proton
projectile will involve two parameters, one characterizing the
momentum degradation and the other flavor flipping.  The
color indices are all averaged over, since multiple gluon
interactions, as the projectile traverses the target nucleus,
are numerous and uncomputable.  Their effects are,
however, quantified in terms of the number, $\nu$, of target
nucleons that participate in the $pA$ collision.  For that
reason it is important that the experimental data must have
centrality selection expressible in terms of the average
$\bar{\nu}$.

In the valon model a proton is considered to consist of three
valons ($UUD$), which have the same flavors as the valence
quarks ($uud$) that they individually contain.  Thus a valon
may be regarded as a parton cluster whose structure can be
probed at high $Q^2$, but the structure of a nucleon itself in
a low-$p_T$ scattering problem is described in terms of the
valons.  As in the parton model we work in a
high-momentum frame so that it is sensible to use the
momentum fractions of the constituents.  Reserving $x$ for
the momentum fraction of a quark, we use $y$ to denote the
momentum fraction of a valon.  In this paper $y$ never
denotes rapidity.  Let the exclusive valon distribution
function be
\begin{eqnarray}
G_{UUD}(y_1, y_2, y_3) = g \, (y_1y_2)^{\alpha}y^{\beta}_3
\, \delta (y_1 + y_2 + y_3 -1) ,
\label{2.1}
\end{eqnarray}
where $y_1$ and $y_2$ refer to the $U$ valons and $y_3$
the $D$ valon. The normalization factor $g$ is
determined by requiring that the probability of finding these
three valons in a proton be one, i.\ e.
\begin{eqnarray}
\int^1_0 dy_1 \int^{1 - y_1}_0dy_2 \int^{1 -
y_1-y_2}_0dy_3 \, G_{UUD}(y_1, y_2, y_3) = 1 .
\label{2.2}
\end{eqnarray}
Denote $\frac{dz}{z}K_{NS}(z)$ the distribution of the valence quark
in a valon, and $\frac{dz}{z}L(z)$ that of the sea quarks, then the distribution
for the favored quark (like $u$ quark in a $U$ valon) in a valon is
$K(z)\equiv K_{NS}(z)+L(z)$. Thus the parton distribution in the proton can be written
as
\begin{eqnarray}
x \,u(x) &= &\iiint dy_1dy_2dy_3G_{UUD}(y_1,y_2,y_3)
\left[2K(x/y_1)+L(x/y_3)\right],\\
x \,d(x) &= &\iiint dy_1dy_2dy_3G_{UUD}(y_1,y_2,y_3)
\left[2L(x/y_1)+K(x/y_3)\right]
\end{eqnarray}
>From previous studies, we use
\begin{eqnarray}
K_{NS} (z) = z^a (1-z)^b/B(a, b + 1),
\end{eqnarray}
\begin{eqnarray}
L (z) = \ell_o (1-z)^5 ,
\end{eqnarray}

$x u(x)$ and $x d(x)$ have been given by CTEQ\cite{cteq4}.
By fitting $x u(x)$ and $x d(x)$ we can get $\alpha,\beta,a,b,\ell_o$
\begin{eqnarray*}
\alpha = 0.70, \quad \beta = 0.25,\quad
a = 0.79, \quad b = -0.26, \quad \ell_0 = 0.083.
\end{eqnarray*}

If we know the probability, $F(x_1, x_2, x_3)$, of finding a $u$ quark at $x_1$, another
$u$ at $x_2$, and a $d$  at $x_3$, and the recombination function, $R_p(x_1, x_2, x_3, x)$,
which is the probability for three quarks to form a proton at $x$, the momentum distribution
of the produced proton in the projectile fragmentation region can be written as
\begin{eqnarray}
\frac{x}{\sigma_{in}}\frac{d\sigma^p}{dx}  \equiv H_p
(x) = \frac{1}{N} \iiint \frac{dx_1}{x_1}\frac{dx_2}{x_2}
\frac{dx_3}{x_3} F(x_1, x_2, x_3)R_p(x_1, x_2, x_3,x)\ ,
\end{eqnarray}
where $N$ is a normalization constant. From former studies
\begin{eqnarray}
R_p (x_1, x_2, x_3, x) = \frac{x_1 x_2 x_3}{
x^3}G_{UUD}\left(\frac{x_1}{x}, \frac{x_2}{x}, \frac{x_3}
{x}\right)
\end{eqnarray}

To calculate $F(x_1, x_2, x_3)$, we need to consider the physics involved
in the passing of the incident proton through the nucleus. During the process,
valons in the proton loss their energy and change their
flavor contents. We describe the energy loss of valons by $D(z_i,\nu_i)$ through the following
equation
\begin{eqnarray}
G'(y'_1, y'_2, y'_3) =\iiint \frac{dy_1 dy_2 dy_3}{y'_1y'_2y'_3}
G(y_1, y_2, y_3)
D(\frac{y'_1}{y_1},\nu_1)D(\frac{y'_2}{y_2},\nu_2)
D(\frac{y'_3}{y_3} \nu_3)
\end{eqnarray}
with $G'(y'_1, y'_2, y'_3)$ depends on $\nu_i$ implicitly. The moments of $D(z_i,\nu_i)$
can be expressed, with a parameter $\kappa$, as
\begin{eqnarray*}
\tilde{D}(n, \nu_i) = \exp \left[\tilde{Q}(n)\nu_i\right]=\exp\left\{- \kappa\nu_i
\left[\psi(n) + \gamma_E\right]\right\}\ .
\end{eqnarray*}

Assume that the flavor changes at each of the $\nu_i$
collisions are incoherent and the probability for a valon
from $U$ to $D$ or from $D$ to $U$ in one valon-nucleon collision is
$q$, then the probability for a valon to change its flavor after $\nu_i$ collisions
can be proved to be $q_{\nu_i}=\left\{1-(1-2q)^{\nu_i}\right\}/2$.
Effectively, one can write valons as
\begin{eqnarray}
U \stackrel{\nu_i}{\rightarrow} p_{\nu_i}U + q_{\nu_i}D,\quad\quad
D \stackrel{\nu_i}{\rightarrow} p_{\nu_i}D + q_{\nu_i}U,
\end{eqnarray}
with $p_{\nu_i}=1-q_{\nu_i}$, then $F(x_1,x_2,x_3)$ can be calculated.
Therefore we can get the momentum distribution of produced proton
in the projectile fragmentation region and compare the results with those
from experiments. In doing so, one needs to calculate $F(x_1,x_2,x_3)$
for the produced protons and anti-protons and make average over all possible combinations of
$\nu_i$, and choose parameters to fit the experimental data on $p-\bar{p}$ production.
We can also calculate the distribution for produced neutrons.
By simultaneously fitting the experimental data on $p-\bar{p}$ and $n-\bar{n}$
in the proton fragmentation region, we get $\kappa=0.62, q=0.37$. This corresponds to
an effective degradation length 17fm. From these two parameters,
one can predict the momentum distribution for the pions in the same region.
In above formulas, we used collision numbers at two different levels: $\nu$ at the nucleon
level, and $\nu_i$ at the valon level.  We need also to take into
account the relation between the number of nucleon-nucleon collisions, $\nu$, and
that of valon-nucleon collisions, $\nu_i$. Because of the space limit, the discussion is not
contained in this talk, but can be found in Ref. [10]. The fitted results and our prediction
can also be found there.

In summary, the general shapes of momentum distribution of hadrons produced in the
proton fragmentation region can be reproduced within the valon model with suitably
chosen parameters and recombination function.

\end{document}